\documentclass{ws-procs975x65}
\font\grb=eurb10
\def\bchi{\hbox{\grb\char'037}\,}

\begin{document}

\title{NEW SOLITON GENERATING TRANSFORMATIONS\\
IN THE BOSONIC SECTOR OF HETEROTIC\\ STRING EFFECTIVE THEORY}

\author{George~A.~ALEKSEEV}

\address{Steklov Mathematical Institute RAS,\\
 Gubkina 8, 119991, Moscow Russia\\
E-mail: G.A.Alekseev@mi.ras.ru}

\begin{abstract}
In the author's paper Ref.~\citen{Alekseev:2009}{}, the integrable structure of the symmetry 
reduced bosonic dynamics in the low energy heterotic string effective theory was presented. 
In that paper, for a complete system of massless bosonic fields which includes metric, 
dilaton field, antisymmetric tensor and any number of Abelian vector gauge fields, considered
in the space-time of $D$ dimensions with $D-2$ commuting isometries, the spectral problem 
equivalent to the symmetry reduced dynamical equations was constructed. However, the 
soliton generating transformations were described in that paper only for the case in which all 
vector gauge fields vanish. In this paper, we recall the integrability structure of these 
equations and describe some new type of soliton generating transformations in which the 
vector gauge fields can also enter the background (seed) solution as well as these can be 
generated even on vacuum background by an appropriate choice of soliton parameters.
\end{abstract}


\bodymatter

\section*{Massless bosonic sector of the low-energy heterotic string theory}

The massless bosonic part of heterotic string effective
action in the string frame is
\[
\label{String_Frame}
{\cal S}=\!\!\displaystyle\int\!\! e^{-\widehat{\Phi}}\!\left\{\widehat{R}{}^{(D)}\!+\nabla_{\scriptscriptstyle{M}} \widehat{\Phi} \nabla^{\scriptscriptstyle{M}}\widehat{\Phi}
-\displaystyle\frac 1{12} H_{\scriptscriptstyle{MNP}} H^{\scriptscriptstyle{MNP}}
-\!\dfrac 12\sum\limits_{\mathfrak{p}=1}^n\displaystyle F_{\scriptscriptstyle{MN}}{}^{(\mathfrak{p})} F^{\scriptscriptstyle{MN}\,(\mathfrak{p})}\!\right\}\!\sqrt{- \widehat{G}}\,d^{D}x
\]
where $M,N,\ldots=1,2,\ldots,D$ and $\mathfrak{p}=1,\ldots n$, ($D$ is the space-time dimension and $n$ is a number of Abelian gauge fields); $\widehat{G}{}_{\scriptscriptstyle{MN}}$ possesses the ``most positive'' Lorentz signature. The components of a three-form $H$ and two-forms $F{}^{(\mathfrak{p})}$ are determined in terms of antisymmetric tensor field $B_{\scriptscriptstyle{MN}}$ and Abelian gauge field potentials $A_{\scriptscriptstyle{M}}{}^{(\mathfrak{p})}$:
\[
H_{\scriptscriptstyle{MNP}}=3\bigl(\partial_{\scriptscriptstyle{[M}} B{}_{\scriptscriptstyle{NP]}}-\!\!\sum\limits_{\mathfrak{p}=1}^n A{}_{\scriptscriptstyle{[M}}{}^{\!(\mathfrak{p})} F_{\scriptscriptstyle{NP]}}{}^{\!(\mathfrak{p})}\bigr),\quad
F_{\scriptscriptstyle{MN}}{}^{\!(\mathfrak{p})}=2\, \partial_{\scriptscriptstyle{[M}} A{}_{\scriptscriptstyle{N]}}{}^{\!(\mathfrak{p})},\quad B_{\scriptscriptstyle{MN}}=-B_{\scriptscriptstyle{NM}}.
\]
Metric $\widehat{G}{}_{\scriptscriptstyle{MN}}$ and dilaton field $\widehat{\Phi}$ are related to the metric  $G_{\scriptscriptstyle{MN}}$ and dilaton  $\Phi$ in the Einstein frame as $\widehat{G}{}_{\scriptscriptstyle{MN}}=e^{2\Phi} G_{\scriptscriptstyle{MN}}$ and $\widehat{\Phi}=(D-2)\Phi$.

\section*{Symmetry reduced bosonic dynamics}

In what follows, we assume that in the space-time of $D$ dimensions with $d=D-2$ commuting Killing vector fields, all "none-dynamical" field components vanish:
\[\left.G_{\scriptscriptstyle{MN}}=\begin{pmatrix}g_{\mu\nu}&0\\
0 & G_{ab}
\end{pmatrix},\hskip1ex
B_{\scriptscriptstyle{MN}}=\begin{pmatrix} 0& 0\\
0 & B_{ab}
\end{pmatrix},\hskip1ex
A_{\scriptscriptstyle{M}}{}^{(\mathfrak{p})}=\begin{pmatrix}0\\ A_a{}^{\!(\mathfrak{p})}\!\!
\end{pmatrix}\hskip1.5ex\right\Vert\hskip1.5ex
\begin{array}{l}
\mu,\nu,\ldots=1,2\\[0ex]
a,b,\ldots=3,4,\ldots D
\end{array}
\]
and all field components and potentials depend only on two coordinates $x^1$ and $x^2$ (one of which can be time-like or both are space-like).
 The coordinates $x^1$, $x^2$ can be chosen so that $g_{\mu\nu}$ takes a conformally flat form $g_{\mu\nu}=f \eta_{\mu\nu}$ with $f(x^1,x^2)>0$. As such coordinates, we choose "geometrically defined" functions $\alpha(x^1,x^2)$ and $\beta(x^1,x^2)$ (generalized Weyl coordinates) and use their linear combinations $\xi$ and $\eta$:
\[\left\{\begin{array}{l} \xi=\beta+j\alpha,\\[1ex]
\eta=\beta-j\alpha,\end{array}\quad\right\Vert\quad
\begin{array}{lccl}\alpha:&&& \det\Vert G_{ab}\Vert=
\epsilon\alpha^2,\\[1.5ex]
\beta:&&&\partial_\mu\beta=\epsilon
\varepsilon_\mu{}^\nu\partial_\nu\alpha,
\end{array}\qquad \eta_{\mu\nu}=\left(\begin{array}{ll}\epsilon_1 & 0\\
0&\epsilon_2\end{array}\right),\hskip1ex
\varepsilon^{\mu\nu}=\begin{pmatrix}0&1\\-1&0\end{pmatrix},
\]
where $\epsilon=-\epsilon_1\epsilon_2$ and $\epsilon_1=\pm1$, $\epsilon_2=\pm1$ are the sign symbols which allow to consider various types of fields. The matrix $\eta^{\mu\nu}$ is inverse to $\eta_{\mu\nu}$ and $\varepsilon_\mu{}^\nu= \eta_{\mu\gamma} \varepsilon^{\gamma\nu}$. The field equations imply that the function $\alpha\ge 0$ is ``harmonic'': $\eta^{\mu\nu}\partial_\mu\partial_\nu\alpha=0$ and
$\beta(x^\mu)$ is defined as its ``harmonically'' conjugated. The parameter $j=1$ for $\epsilon=1$ (the hyperbolic case) and $j=i$ for $\epsilon=-1$ (the elliptic case). Therefore the coordinates $\xi$ and $\eta$ respectively both are real or complex conjugated to each other.

\section*{The spectral problem equivalent to dynamical equations}
As it was shown in Ref.~\citen{Alekseev:2009}{}, the symmetry reduced dynamics of massless bosonic fields in heterotic string effective theory  is integrable and the solution of the dynamical equations is equivalent to solution of the spectral problem found there. This spectral problem is formulated in terms of four $(2 d+n)\times(2 d+n)$-matrices $\mathbf{\Psi}(\xi,\eta,w)$, $\mathbf{U}(\xi,\eta)$, $\mathbf{V}(\xi,\eta)$, $\mathbf{W}(\xi,\eta,w)$ which should satisfy the linear system for $\mathbf{\Psi}$ with the algebraic conditions which determine the canonical Jordan forms of its coefficients:
\begin{equation}\label{LinSys}
\left\{\begin{array}{l}
2(w-\xi)\partial_\xi \mathbf{\Psi}=\mathbf{U}(\xi,\eta) \mathbf{\Psi}\\[2ex]
2(w-\eta)\partial_\eta \mathbf{\Psi}=\mathbf{V}(\xi,\eta) \mathbf{\Psi}
\end{array}\quad\right\Vert
\qquad\begin{array}{l}
{\bf U}\cdot{\bf U} ={\bf U},\hskip3ex \text{tr}{\bf U}=d, \\[2ex]
{\bf V}\cdot{\bf V} ={\bf V},\hskip3ex\text{tr}{\bf V}=d,
\end{array}
\end{equation}
and this system should admit a symmetric matrix integral $\mathbf{K}(w)$ such that
\begin{equation}\label{W_Condition}
\left\{\begin{array}{l}
{\bf \Psi}^T {\bf W} {\bf \Psi}={\bf K}(w)\\[2ex]
{\bf K}^T(w)={\bf K}(w)
\end{array}
\qquad\right\Vert\qquad
\dfrac{\partial\mathbf{W}}{\partial w}=\mathbf{\Omega},\qquad\hskip1ex
\mathbf{\Omega}=\begin{pmatrix}
0&I_d&0\\I_d&0& 0\\0&0&0
\end{pmatrix}
\end{equation}
where $w\in \mathbb{C}$ is a spectral parameter, $\mathbf{\Omega}$ is $(2 d+n)\times (2 d+n)$-matrix,  $I_d$ is a $d\times d$ unit matrix. Denoting by $\mathbf{W}_{(3)(3)}$  the lower right $n\times n$ block of $\mathbf{W}$, we require also
\begin{equation}\label{reality}
\overline{\mathbf{\Psi}(\xi,\eta,w)}= \mathbf{\Psi}(\overline{\xi},\overline{\eta},\overline{w}),\qquad
\overline{\mathbf{K}(w)}= \mathbf{K}(\overline{w}),\qquad
\mathbf{W}_{(3)(3)}=I_n.
\end{equation}
In accordance with Ref.~\citen{Alekseev:2009}{}, any solution $\{\mathbf{\Psi},\mathbf{U},\mathbf{V},\mathbf{W}\}$ of the the Eqs. (\ref{LinSys})--(\ref{reality}) determines uniquely some solution of the dynamical equations and vice versa.

\section*{Soliton generating transformation}
Given some solution as background for solitons, we denote its $(2 d+n)\times (2 d+n)$-matrices by "$\circ$", and for one soliton on this background we assume
\[
\mathbf{\Psi}=\bchi\cdot {\overset \circ {\mathbf{\Psi}}},\qquad \bchi=\mathbf{I}+\dfrac {\mathbf{R}(\xi,\eta)}{w-w_o}, \qquad \bchi^{-1}=\mathbf{I}+\dfrac {\mathbf{S}(\xi,\eta)}{w-w_o}\quad\Longrightarrow\quad \det \bchi\equiv 1
\]
where $w_o$ is a real constant and  $(2 d+n)\times (2 d+n)$-matrices $\mathbf{R}$ and $\mathbf{S}$ are real and depend on $\xi$, $\eta$ only. We also assume
$\mathbf{K}(w)={\overset \circ {\mathbf{K}}}(w)$. Then the consistency conditions $\bchi \bchi^{-1}=\bchi^{-1} \bchi=\mathbf{I}$ imply  $\mathbf{S}=-\mathbf{R}$ and $\mathbf{R}\cdot\mathbf{R}=0$.
This means that $\mathbf{R}$ is degenerate and for simplicity, we consider $\mathbf{R}$ having the rank equal to $1$:
\[\mathbf{R}=\mathbf{n}\otimes\mathbf{m},\qquad (\mathbf{m}\cdot\mathbf{n})=0,
\]
where $\mathbf{m}(\xi,\eta)$ and $\mathbf{n}(\xi,\eta)$ are $(2 d+n)$-vector row and column respectively.
Substitution of the above expressions into the Eqs. (\ref{LinSys})--(\ref{reality}) leads to a set of relations at the pole $w=w_o$ and at $w\to\infty$ which can be solved explicitly. Thus we obtain:
\[\mathbf{U}={\overset\circ {\mathbf{U}}}+2\,\partial_\xi\mathbf{R},\qquad
\mathbf{V}={\overset\circ {\mathbf{V}}}+2\,\partial_\eta\mathbf{R},\qquad
\mathbf{W}={\overset\circ {\mathbf{W}}}-\mathbf{\Omega}\cdot \mathbf{R}-\mathbf{R}^T\cdot\mathbf{\Omega}
\]
The vector functions $\mathbf{m}(\xi,\eta)$ and $\mathbf{n}(\xi,\eta)$ are determined by the expressions
\[\mathbf{n}=\ae^{-1}\,\,\mathbf{p},\qquad
\mathbf{m}=\mathbf{k}\cdot{\overset\circ {\mathbf{\Psi}}}{}^{-1} (\xi,\eta,w_o),\qquad
\mathbf{p}={\overset\circ {\mathbf{\Psi}}} (\xi,\eta,w_o)\cdot\mathbf{l},\qquad
\ae=\dfrac 12(\mathbf{p}^T\cdot\mathbf{\Omega}\cdot\mathbf{p}).
\]
where the real $(2 d+n)$-vectors $\mathbf{k}$ and $\mathbf{l}$ are constant; $\mathbf{k}$ is determined unequally in terms of $\mathbf{l}$, and the choice of $\mathbf{l}$ is arbitrary provided it satisfies an algebraic constraint:
\[\mathbf{k}=\mathbf{l}^T\cdot{\overset\circ{\mathbf{K}}}(w_o),\qquad
(\mathbf{l}^T\cdot{\overset\circ{\mathbf{K}}}{}(w_o)\cdot \mathbf{l})=0.
\]
Thus, choosing any background solution and any real constant $(2 d+n)$-vector column $\mathbf{l}$ which satisfy only the last mentioned constraint, we can construct the matrix $\mathbf{W}$ whose components (in accordance with the expressions given in Ref.~\citen{Alekseev:2009}) allow us to calculate all field components and potentials of generated one soliton solution.

It is necessary to mention, however, that for stationary case, generation of one soliton can lead (similarly to vacuum solitons of Belinski and Zakharov) to a change of signature of metric and therefore the number of solitons in this case should be even. On the other hand, the described here one soliton generating transformation  can be generalized easily to the multisoliton case.

\section*{Acknowledgements}
The author is thankful to the Organizing Committee of
MG12 for partial financial support for participation in the Meeting. This work was supported in parts by the Russian Foundation for Basic Research (grants 08-01-00501, 08-01-00618, 09-01-92433-CE) and the program ``Mathematical Methods of Nonlinear Dynamics'' of the Russian Academy of Sciences.

\end{document}